\documentstyle[preprint,aps,eqsecnum]{revtex}

\begin{document}

\draft
\title{  Quasiperiodic Modulated-Spring Model} 
\author{H. Hiramoto$^1$ and Mahito Kohmoto$^2$}
\address{$^1$Department of Physics, College of Humanities and
Sciences, Nihon University, Sakurajosui, Setagaya-ku, Tokyo 156, Japan} 
\address{$^2$Institute for Solid State Physics, 
University of Tokyo, Roppongi, Minato-ku, Tokyo 106, Japan}

\maketitle

\begin{abstract}

We study the classical vibration problem of a chain with spring constants
which are 
modulated in a quasiperiodic manner, {\it i. e.}, a model in which the elastic
energy  is $\sum_j k_j( u_{j-1}-{u_j})^2$, where $k_j=1+\Delta
cos[2\pi\sigma(j-1/2)+\theta]$
 and $\sigma$ is an irrational number. 
For $\Delta < 1$, it is shown analytically that the spectrum is  absolutely
continuous, {\it i.e.},  all the
eigen modes are extended. For $\Delta=1$, numerical scaling analysis shows
that  the spectrum is purely singular continuous, {\it i.e.}, all the modes
are critical.  
\end{abstract}

\narrowtext
\newpage

\section {Introduction}

For some years, the electronic properties of quasiperiodic
systems  have been extensively studied. In particular, a number of studies
are devoted to a one-dimensional quasiperiodic tight-binding model \cite{1},

\begin{equation}
   -\psi_{n+1} -  \psi _{n-1} + \lambda V(n\sigma )\psi_n = E\psi_n, 
\label{qpTB} 
\end{equation}
where $V(x)$ is a periodic function with period 1 ( $V(x+1)=V(x)$ ) and
$\sigma$ 
is an irrational number.

The properties of the energy spectra and the eigen states of (\ref{qpTB}) are 
generically as follows\cite{2,3}:  When the strength of the  potential 
$\lambda$
is small, all the states are extended, {\it i.e.}, the energy spectrum is
purely absolutely continuous. As $\lambda$ is increased, localized
states appear, which are separated from extended states by
mobility edges in the spectrum. That is, the spectrum has absolutely continuous
parts and pure point parts, and they are separated by mobility edges. When
$\lambda$ is large enough, all the states are localized, {\it i.e.}, the
spectrum is pure dense point.

The well-known exceptions 
are the Harper model \cite{4} and the Fibonacci model \cite{5,6}. 
In the Harper model ($V(x)=cos(2\pi x) $), all the states are extended
for $\lambda <  2$, while all the states are localized for $\lambda >2$. 
At $\lambda=2$, all the states are critical and
the energy spectrum is purely singular continuous. These  properties are
caused by the existence
of  the duality \cite{4}. It is self-dual at the critical point
$\lambda=2$.

In the Fibonacci model $V$'s take two values $1$ and $-1$ arranged by
the Fibonacci sequence. So $V(x)$ is piecewise constant. The spectrum is purely
singular continuous and all the states are critical  irrespective
of  $\lambda$. 

In this paper, we study the dynamics of a one-dimensional array of equal mass
particles  connected by springs modulated in a quasiperiodic manner. 
The model called the modulated spring (MS) model is introduced in Sec. 2. A
quasiperiodic case was first 
studied by De Lange and Janssen \cite{7}. Their numerical results show the
intricate Cantor-set like structures of the spectra. Janssen and Kohmoto
\cite{8}
applied the multifractal method \cite{9,10} and conjectured that the model has a
pure singular continuous spectrum at $\Delta=1$ and has a mixed one for
$\Delta< 1$.
However, their system sizes  are not large enough. In fact, their
claim for $\Delta <  1$ is not correct as shown below. 

Incidentally the MS model is related to  a two-dimensional tight-binding
model in
a  magnetic field with next-nearest-neighbor hoppings \cite{11,13,13p} which is
introduced in Sec. 3. Using the  result of Thouless \cite{11} for the Lyapunov
exponent it is shown that all the eigen modes are extended for $\Delta < 1$
in Sec.4.  
For $\Delta = 1$ numerical
scaling analyses are required to determine the types of the spectrum. 
The results are presented in Sec.5 and it is
concluded  conclude that all the eigen modes are critical for $\Delta = 1$.   
This is presumably due to
the fact that the spring constants can be infinitesimally close to zero.

\section {Modulated spring (MS) model}

The equations of motion of the one-dimensional MS model are 

\begin{equation}       
   \omega^2 u_j = k_j(u_j -u_{j -1})+k_{j+1}(u_j -u_{j +1})   \label{eqmot}
\end{equation}  
where $u$'s are  displacements of the  atoms and $k$'s are the 
spring constants:                                                 

\begin{equation}
k_j=1+\Delta
cos[2\pi\sigma(j-1/2)+\theta].                                \label{spring} 
\end{equation}
Note that $\Delta$ cannot
be greater than one since the spring constants $k$'s must be positive. Thus
$0<\Delta \leq 1$.  The MS model can be regarded as   a one-dimensional
electronic 
tight-binding model by rewriting (\ref{eqmot}) with (\ref{spring}) as 

\begin{equation} 
-k_j u_{j-1} - k_{j+1} u_{j+1} +V_ju_j=Eu_j, \label{vibration}
\end{equation}
with
  
\begin{equation} V_j =k_j
+k_{j+1} -2 = 2 \Delta cos(\pi \sigma) cos(2\pi\sigma j+\theta),  \label{V} 
\end{equation}     
and

\begin{equation}
E= \omega^2 -2.                                                  \label{energy}
\end{equation}

\section {tight-binding electrons in a magnetic field on the square
lattice with next-nearest-neighbor hopping }

Consider the tight-binding model on the square lattice with  nearest-neighbor
hoppings $t_a$ and $t_b$,  next- nearest-neighbor hopping  $t_c$, and  magnetic
flux  per plaquette  $\phi$  (see Fig. 1). In the Landau  gauge ($\vec{A} = (0,
By)$ ) the tight-binding equation is 

\begin{eqnarray}
&&-t_a (\psi_{m-1,n} + \psi_{m+1,n}) -t_b (e^{i2\phi \psi m}\psi_{m,n-1}
e^{-i2\pi \phi m} \psi_{m,n+1})  \nonumber \\
&&-t_c (e^{-i2\pi \phi (m-1/2)} \psi_{m+1,n+1} +e^{i2\pi \phi (m+1/2)}
\psi_{m-1,n-1} \nonumber \\ 
&&e^{i2\pi \phi (m+1/2} \psi_{m+1,n-1} + e^{-i2\pi \phi
(m-1/2} \psi_{m+1,n+1})    
 = \; E \psi_{m,n}. \label {scheq} 
\end{eqnarray} 
Since this equation is
translation invariant in the y-direction, the Bloch theorem is applied  and
one may write 

\begin{equation}
\psi_{m,n} = e^{-ikn} \varphi_m \;.                       \label{Bloch}
\end{equation}  
By substituting (\ref {Bloch}), (\ref {scheq})  reduces to a
one-dimensional tight-binding equation

\begin{eqnarray}    
&&-(t_a+2t_c cos[k+ 2\pi \phi (m-1/2)] \varphi_{m-1} - (t_a + 2t_c cos[(k+ 2\pi
 \phi (m+1/2)]) \varphi_{m+1} \nonumber \\
&&\qquad\qquad\qquad-2t_b cos(k+2 \pi \phi m) \varphi_m \;= \; E\varphi_m
\;.  \label {TB} 
\end{eqnarray}

Since the original 2d problem  has a symmetry with respect
 to an interchange of  $x$ and $y$ (\ref{TB}) has a duality. Substitute
\begin{equation}
\varphi_m = \sum_j f_j e^{i\left[Km+(2\pi \phi m-k)j\right]},      
\label{varphi}   
\end{equation} 
 to (\ref{TB}), then one gets

\begin{eqnarray}
&&-(t_b+2t_c cos\left[K+ 2\pi \phi (j-1/2)\right] f_{j-1} - 
(t_b + 2t_c cos\left[(K+ 2\pi
 \phi (j+1/2)\right]) f_{j+1} \nonumber \\
&&\qquad\qquad\qquad-2t_a cos(K+2 \pi \phi j) f_j \;= \; Ef_j \; 
\label{TBdual}  
\end{eqnarray}
which has the same form as (\ref{TB}).

\section {Lyapunov exponent}

Using the duality in the last Sec.3 Thouless \cite{11} obtained the Lyapunov
exponent
$\gamma$ ( inverse of the localization length) for (\ref{TB}) as

\begin{eqnarray}
\gamma&=& 
\left\{
\begin{array}{cc}
 ln(t_b + \sqrt{t_b^2 - 4t_c^2}) - ln(t_a + \sqrt{t_a^2 - 4t_c^2})
 & {\rm for}\; 2t_c < t_a < t_b,\\
ln(t_b + \sqrt{t_b^2 - 4t_c^2}) - ln(2t_c) & {\rm for}\;t_a < 2t_c < t_b,\\  
0 & {\rm for}\;t_a < t_b < 2t_c.
\end{array}
\right.                                     \label{gamma}
\end{eqnarray}

 From (\ref{gamma})  the Lyapunov exponent is strictly
positive for $2t_c < t_b$ and $t_a < t_b$. Thus all the eigen states of (\ref{TB})
are exponentially localized and the spectrum  is pure point. The duality
(\ref{varphi}) which maps  (\ref{TB}) to (\ref{TBdual}) by exchanging $t_a$ and
$t_b$  is essentially  a Fourier transformation. So a localized state is mapped
 to an extended state. Therefore all the states are extended for if 

\begin{equation}
2t_c < t_b \;{\rm and}\; t_a < t_b     \label{tab}
\end{equation}
Note that the MS model (\ref{vibration}) with (\ref{V}) and 
(\ref{TB}) are the same if
\begin{equation}
t_a=1,\; t_c=\Delta/2,\; t_b= \Delta cos(\pi\sigma) {\rm and}\; k=\theta + \pi.
\label{equi} 
\end{equation}
For $\Delta < 1$, (\ref{tab}) is satisfied.
This means that {\it all the eigen modes of the MS model (\ref{eqmot}) is
extended
for $\Delta < 1$.}

For $\Delta = 1$ we have
$2t_c =t_a$ and  $t_b < t_a$. In the dual situation  ($2t_c =t_b ,t_a <
t_b$) one
has $\gamma=0$ from (\ref{gamma}) and the states are either critical or
extended.
Thus in the original situation  the eigen modes are either critical or
localized.  In order to determine whether the eigen modes are  critical or
localized for $\Delta = 1$
numerical  analysis is required.It is presented in the next section.

\section {Numerical Scaling analysis}

We take $\sigma$ to be $(\sqrt{5}-1)/2$ 
(the inverse
of the golden mean) and $\theta$ to be 0. This irrational number $\sigma$ is
approached by rational numbers $F_{n-1}/F_n$ ($F_{n-1}/F_n \rightarrow 
\sigma$ as
$n  \rightarrow \infty$),  where $F_n$ is
the $n$th Fibonacci number defined by $F_n=F_{n-1}+F_{n-2}$ and $F_0=F_1=1$.
Thus in the numerical calculations $\sigma$ is approximated by $F_{n-1}/F_n$
which makes the system  periodic with period $F_n$. The spectrum consists
of $F_n$ bands. As $n$ is increased, {\it i.e.}, as $F_{n-1}/F_n$ approaches to
$\sigma$, each band splits into sub-bands and the width of each band
approaches to zero. 
We analyze scaling behaviors of
the band widths for large $n$'s. For the band width, we adopt the width of
$\omega^2$  
or $E$ in (\ref{energy}) and
denote it as $\delta(\omega^2)$.

For localized  modes the band widths are expected to decrease
exponentially with $F_n$, while for  extended modes they should behave as
$1/F_n$ \cite{kohmoto}. Thus, if we define a scaling exponent $\alpha$ by 
$1/F_n \sim \delta(\omega^2)^\alpha$,  eigen modes are localized if 
$\alpha$ is 0 and
are extended if $\alpha$ is 1. In addition, extended modes with $\alpha=1/2$ may
exist. These correspond to the remnant of van Hove singularities at the edges
of band clusters. This behavior was observed in the Harper model in the extended
region \cite{12}. For critical modes, which correspond to a singular continuous
spectrum, $\alpha$ takes a continuous range of values.  
As $n$ is increased by $2$ or $3$, each band splits into three subbands.
Therefore,
each point in the spectrum of the quasiperiodic limit ($n \rightarrow \infty$ )
is specified by an infinite series of 1, 0 and -1, which represents the upper,
middle and lower sub-bands, respectively. In Fig.1 $F_n\delta (\omega^2)$ is
plotted against $n$ for several points in the spectrum for (a) $\Delta
=0.95$ and
(b) $\Delta=1.0$ \cite{D1}. 
For $\Delta=0.95$,  the widths scale as
$\delta(\omega^2)\sim   1/F_n$ except for the points identified by
$\{C_1 C_2 \cdots   -1 -1 -1 \cdots \}\;$ or 
$\{C_1 C_2 \cdots    1\;1\;1\;\cdots \}$
where they scale as $\delta(\omega^2)\sim (1/F_n)^2$. The latter scaling with
$\alpha = 1/2$ are  remnants of the van Hove singularities. The spectrum is
absolutely continuous, {\it i.e.}, the modes are extended at least for 
those in Fig. 2(a).

For $\Delta=1, \delta(\omega^2)$ scales with various powers of $1/F_n$ as shown
in Fig. 2(b). It is an interesting result that the lowest edge band $\{ -1 -1 -1
-1 -1\;\cdots  \}$ seems to scale exactly as $\delta(\omega^2) \sim (1/F_n)^3$. 
The scaling  $\alpha = 1/3$ is different from the edge scaling of absolutely
continuous spectra $\alpha = 1/2$ which represents a remnant of a van Hove
singularity.  Thus it seems  that usual extended modes do not exist in the
vicinity of $\omega=0$.

In order to clarify the global properties of the spectrum, we perform
 the multifractal analysis \cite{9,10}. We calculate $f(\alpha)$ of the
whole spectrum which is defined by $\Omega(\alpha)d\alpha \sim
F_n^{f(\alpha)}$, 
where $\Omega(\alpha)d\alpha$ is the the number of bands whose scaling
exponents  $\alpha$'s lie in an interval [$\alpha$, $\alpha+d\alpha$].
When we calculate $f(\alpha)$ using the statistical mechanics-like formalism
\cite{10} a smooth $f(\alpha)$
on $\left[\alpha_{min}, \alpha_{max}\right]$ is always obtained for a
finite system
and it is not possible to distinguish spectral types. Therefore it is 
crucial to
take an extrapolation  $n \rightarrow \infty$.

In Fig. 3,  (a) $\alpha_{max}^{(n)}$ and (b) $f_{max}^{(n)}$ (calculated with
$\sigma_n =  F_{n-1}/F_n$) for $\Delta=0.95$ are plotted. It is clearly
seen that
$\alpha_{max}= 1$ and $f(\alpha_{max}) =1$ in the $n \rightarrow \infty$ limit.
Similar results are obtained for other values of $\Delta$ which are less than
one.  These numerical  results are consistent with the analytical result in
Sec. 4
that the spectrum is absolutely continuous  for $\Delta <1$.

In Fig. 4, $f(\alpha)$ which is obtained  by extrapolation  $n \rightarrow
\infty$
 for
$\Delta=1$ is shown. This gives the convincing evidence that the spectrum
is purely
singular continuous rather than point like. Thus we conclude that
{\it all the eigen modes of the MS spring model (\ref{eqmot}) is critical at
$\Delta=1$}.

\begin {acknowledgements}
  It is a pleasure to thank D. J. Thouless for very useful discussions.
\end {acknowledgements}


\begin{references}
\bibitem{D1} Note that, when $\Delta=1$, $k$'s for certain $j$'s
become exactly zero if n=3m+1 (m: integer). Since the infinitely long chain is
divided  into finite chains in this case all the
modes are  trivially localized and $\delta$'s are zero. This is caused by
the choice $\theta = 0$ in (\ref{spring}) and is not a generic feature. Thus
we disregard the data for $n=3m+1$ when $\Delta=1.0$. 



\figure{1}  The square lattice with nearest- neighbor and next- nearest-
neighbor hoppings in the magnetic field $\vec B$ whose flux quanta per 
plaquette is $\phi$.  
                                
\figure{2}  $F_n\delta (\omega)$ against $n$ for several modes with (a) 
$\Delta=0.95$ and (b) $\Delta= 1.0$.

\figure{3} (a) $\alpha_{max}(n)$ and (b) $f_{max}(n)$ against $1/n$ for
$\Delta=  0.95$.
 
\figure{4}  $f(\alpha)$ for spectrum with
$\Delta=1$.
\end{references}
\end{document}